# Extreme plasma states in laser-governed vacuum breakdown

Evgeny S. Efimenko[1,*], Aleksei V. Bashinov[1], Sergei I. Bastrakov[2], Arkady A. Gonoskov[1,2,3], Alexander A. Muraviev[1], Iosif B. Meyerov[2], Arkady V. Kim[1] and Alexander M. Sergeev[1]

[1]Institute of Applied Physics, Russian Academy of Sciences, Nizhny Novgorod 603950, Russia
[2]Lobachevsky State University of Nizhni Novgorod, Nizhny Novgorod 603950, Russia
[3]Department of Physics, Chalmers University of Technology, SE-41296 Gothenburg, Sweden

**Triggering vacuum breakdown at the upcoming laser facilities can provide rapid electron-positron pair production for studies in laboratory astrophysics and fundamental physics. However, the density of the emerging plasma should seemingly stop rising at the relativistic critical density, when the plasma becomes opaque. Here we identify the opportunity of breaking this limit using optimal beam configuration of petawatt-class lasers. Tightly focused laser fields allow plasma generation in a small focal volume much less than $\lambda^3$, and creating extreme plasma states in terms of density and produced currents. These states can be regarded as a new object of nonlinear plasma physics. Using 3D QED-PIC simulations we demonstrate the possibility of reaching densities of more than $10^{25}$ cm$^{-3}$, which is an order of magnitude higher than previously expected. Controlling the process via the initial target parameters gives the opportunity to reach the discovered plasma states at the upcoming laser facilities.**

In the nearest years several upcoming laser-facilities (1; 2; 3; 4) are expected to trigger the cascaded production of electron-positron pairs through the processes of quantum electrodynamics (QED) (5; 6; 7). In contrast to early experiments of Burke et al. (8), where pre-accelerated electrons have been used, in this objective the laser fields are used for not only triggering the generation of hard photons and their decay into pairs but also for accelerating the produced particles so that the field energy feeds the avalanche of pair production. Although this type of cascade requires significantly weaker laser fields than the Schwinger field (9; 10; 11; 12; 13; 14), the growth rate (15) can be sufficiently high to enable extreme scenarios that are remarkable in many respects, ranging from general questions of relativistic plasma dynamics to generation of high fluxes of gamma photons and high-energy particles (16; 17; 18; 19; 20; 21; 22; 23; 24; 25). Despite the progress in QED-PIC simulations (26; 27; 28; 29), most of the theoretical studies have been carried out for the QED-assisted particle dynamics in a given field of various configurations (30; 31; 26; 32; 33; 34; 35; 36; 37; 38) (39; 40; 41), whereas the self-action of the emerging electron-positron plasma (14; 22) and its ultimate states to a large degree remain yet undiscovered.

In the present study we consider reaching such an ultimate state of plasma during cascade development in the e-dipole field (42) which is the utmost case of the colliding beam concept (9; 13; 11). Apart from reaching the strongest possible field for a given power of a laser facility, the e-dipole wave provides a remarkably localized field in a volume of much less than $\lambda^3$ and enhances particle localization in a small bulk around the centre by the anomalous radiative trapping effect (43). We show that with the density growth in this bulk the extreme alternating currents driven by the laser field make the plasma unstable and lead to plasma stratification into thin sheets, where the plasma reaches an extreme state in terms of current and density. At a later stage they merge and eventually form two aligned sheets. We provide thoughtful analysis and explanation of the underlying physics and use 3D QED-PIC simulations (29) to demonstrate the possibility of reaching this scenario at the upcoming laser facilities with a total peak power of around 10 PW.

**Vacuum breakdown in an e-dipole wave**

We start with showing how the extreme states of electron-positron plasma emerge in a particular case of vacuum breakdown governed by an e-dipole wave with a constant total power of 10 PW raised smoothly over one cycle. The result of a 3D QED-PIC simulation of this process is shown in Fig. 1.

Once the e-dipole wave power exceeds $P_{th}$ = 7.2 PW rapid pair production enables exponential growth of the number of particles in the small vicinity of the center (see Methods section Threshold of the breakdown). While plasma density is well below the critical value, the back reaction of the plasma onto the field is negligible, and a QED cascade develops in the given fields. This **linear stage** is well studied for various field structures, such as plane standing waves (44; 15; 10; 26), several focused laser pulses (11; 9) and an e-dipole wave (17). An e-dipole wave has a standing-wave-like structure where particle motion is strongly affected by the radiation losses leading to the anomalous radiative trapping (ART) effect (43; 37). In the ART regime the particles oscillate predominantly along the electric field vector gaining energy up to $\gamma \sim a_0$, here $\gamma$ is the Lorentz factor and $a_0$ is the dimensionless amplitude of the electric field. The particles form a narrow cylindrical column in the center with the height of ~ $0.5\lambda$ along the $z$ axis and the radius of ~ $0.2\lambda$ (see Fig. 1(e)), where $\lambda$ = 910 nm is the wavelength of laser radiation forming the dipole wave.

The **nonlinear stage** occurs when the pair density exceeds $10^{23}$cm$^{-3}$ and becomes comparable with the relativistic critical value $\gamma n_c$, where $n_c = \pi m c^2/(\lambda e)^2$ is the critical plasma density for the laser radiation, $c$ is the light velocity, $e, m$ are electron charge and mass, respectively. At this stage, dense electron-positron pair plasma starts to influence the field structure leading to a gradual decrease of the field amplitude, while the field structure remains almost non-perturbed. Lower field amplitude implies lower pair growth rate, which manifests itself as a gradual slope decrease of the curve describing time evolution of positron density $n_p(t)$ at *11-15T* in Fig. 1(b).

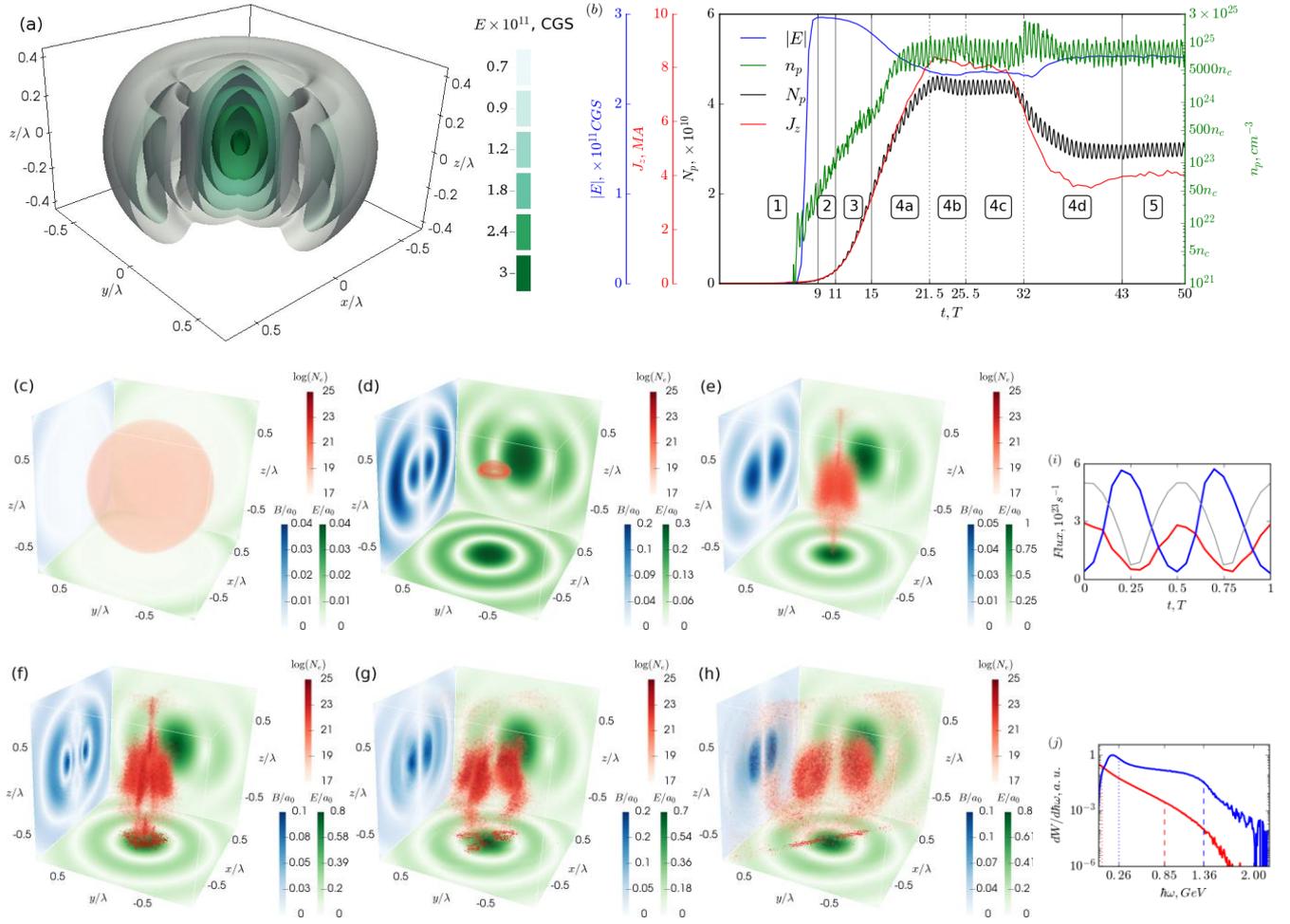

**Figure 1. 3D PIC simulation of vacuum breakdown in a 10 PW e-dipole wave.** (a) Contour plot of the electric field of the e-dipole wave. (b) Timeline of interaction. Main stages are shown: 1 - target compression and formation of a standing wave structure; 2 - linear cascade; 3 - nonlinear stage; 4 – development of current instability: 4a – linear stage of current instability and merger to four sheets, 4b,c – merger to three and two sheets, respectively, 4d – relaxation to a stationary state; 5 - final stationary state. The curves depict maximum electric field $E$ (blue solid line), maximum positron density $n_p$ (green solid line), total current $J_z$ in z direction through plane z=0 (red line), number of positrons in a cylinder with diameter and height of λ $N_p$ (black line), stationary value of electric field equal to $2.51\times10^{11}$ CGS (blue dashed line), stationary value of positron density averaged over the period equal to $7\times10^{24}$ cm$^{-3}$(green dashed line). (c-h) Plasma-field structure evolution: (c) initial plasma distribution; (d) compression of the target; (e) linear stage of cascade; (f-g) current instability development; (h) final stationary state. Green and blue surface depict magnetic $B$ and electric $E$ fields respectively. Electron density $N_e$ is plotted to a logarithmic scale is shown by red. Red contour at the bottom plane depicts plasma contour at plane z=0. (i) Temporal structure of emitted positrons (blue), photons (red) with energy exceeding 1 GeV and electric field in the center (gray) in the stationary state. (j) Spectra of emitted positrons (blue), photons (red) in the stationary state averaged over the wave period, dashed lines show the maximum energy such that particles with energy exceeding this value possess 1% of total particle energy, dotted lines depict average particle energy**.**

When growth rate becomes comparable to relativistic plasma frequency the sudden effect of symmetry breaking of the axisymmetric laser-plasma distribution comes into play due to **current instability**. The quasistationary plasma-field structure starts to be unstable along the azimuthal angle φ, which leads to the development of small-scaled density perturbations. Since electrons and positrons moving in opposite directions produce strong axial current, its azimuthal fluctuations generate an alternating radial magnetic field $B_\rho$. The pairs begin to attract to $B_\rho$ nodes thereby locally increasing current density, which, in turn, causes additional growth of $B_\rho$, consequent stronger compression of pairs in the vicinity of the nodes and further growth of current density (see Methods section Symmetry breaking). Initially, the pair plasma column breaks down into many sheets, which in the plane z=0 look like narrow rays. This process manifests itself as a rapid increase of $n_p(t)$ slope close to $t=15T$ in Fig. 1 (b). Later, due to current interactions, the sheets are attracted to each other forming denser sheets with correspondingly higher currents. After each merger system relaxes to the new equilibrium state with lower total current and number of particles. Each merger can be distinguished as a local peak in the positron density curve in Fig. 1(b), mergers into four ($t=21.5T$), three ($t=25.5T$) and two ($t=32T$) current sheets are shown in Fig. 1 (f,g,h). This effect is the most prominent for the last merger to the final **steady state** in the form of two aligned sheets. The sheets form a single plane oriented at a random azimuthal angle. This plasma-field structure is stable because it preserves the minimum axial symmetry and consists of the minimum number of sheets distanced from each other along the azimuth as much as possible while the electric field hump confines radial particle motion.

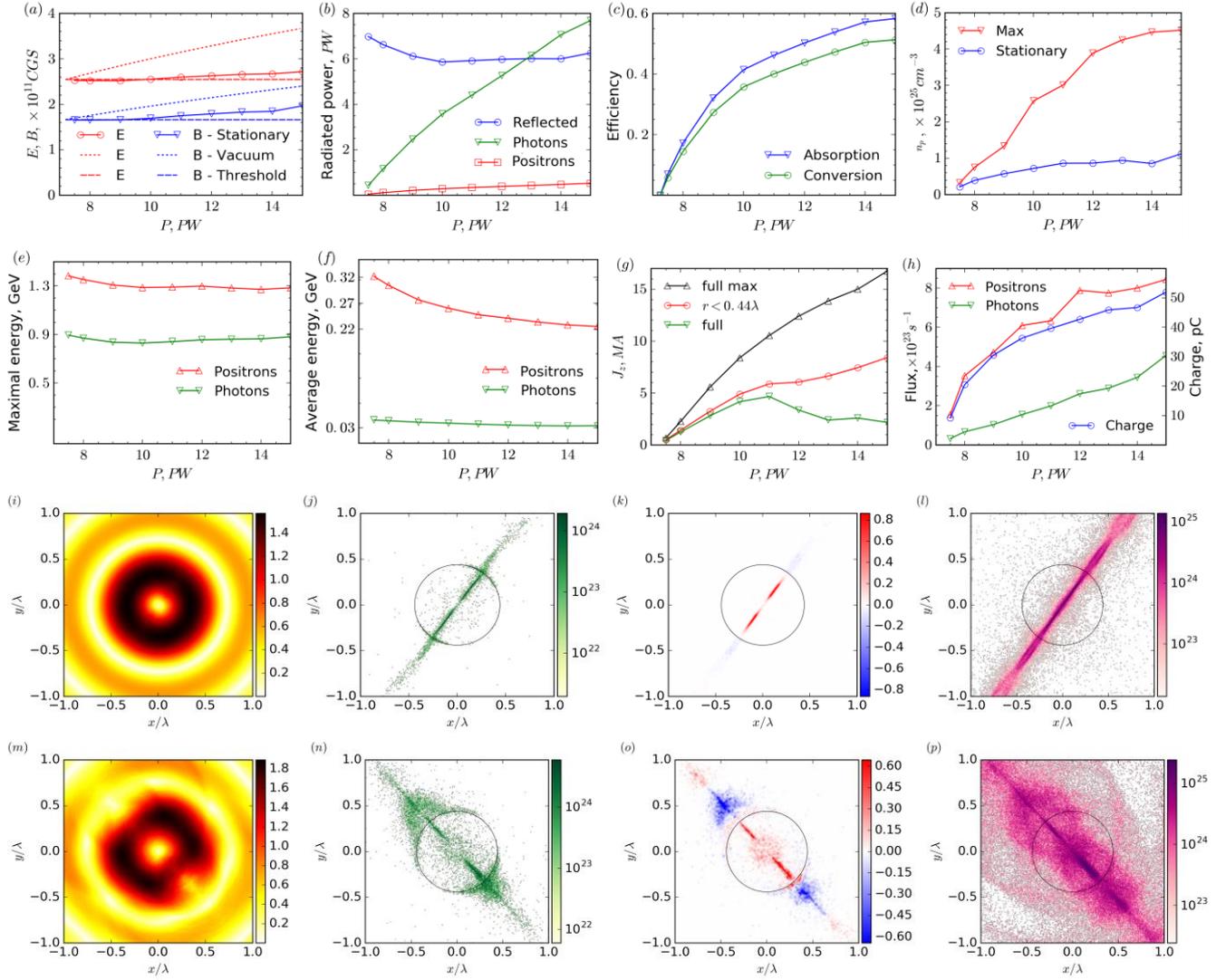

**Figure 2. Properties of the stationary state *vs* E-dipole wave power**. (a) Maximum electric (solid red line) and magnetic (solid blue line) field in stationary state. Dotted lines represent vacuum fields scaling as $P^{1/2}$, dashed lines show electric and magnetic field values corresponding to $P_{th}$. (b) Energy balance in stationary state: reflected energy (blue line), energy transferred to photons (green line) and positrons (red line). (c) Efficiency of energy absorption (blue line) and conversion to gamma photons (green line). (d) Pair plasma density in stationary state (blue line) and maximal value on transient nonlinear stage (red line). (e) Maximum energy of positrons (red line) and photons (green line). (f) Average energy of positrons (red line) and photons (green line). (g) Current $J_z$ through plane $z=0$ in stationary state: total current (green line), current in the central antinode region (radius $r < 0.44\lambda$) (red line) and maximum current during all stages (black line). (h) Total charge of the single positron burst (blue line), averaged over period electron (red line) and photon (green line) flux with energy exceeding 1 GeV. (i-p) Structure of positron density and currents in stationary state at $z=0$ plane for (i-l) 8 PW and (m-p) 15 PW. From left to right: (i,m) magnetic field ($\times 10^{11}$ CGS), (j,n) positron density(cm$^{-3}$) plotted to a logarithmic scale, (k,o) current density($\times 10^{16}$ A/cm$^2$), (l,p) photon density(cm$^{-3}$) plotted to a logarithmic scale.

## Self-consistent plasma-field states

Self-consistent plasma-field states are the result of nonlinear interaction of the incident wave with the self-generated pair plasma through QED cascades which are dependent on laser wave power. These states are stationary on the average over the laser cycle, i.e. the average growth rate is zero and the total number of particles does not change. Nevertheless, within the laser period the density changes and particle escape from the focal region is compensated by production in the intense field. Structure of steady state fields is close to a dipole wave, so, according to notion of a vacuum breakdown threshold (see Methods section Threshold of the breakdown), steady state field amplitude has to be close to the threshold value. Simulations show, that for power below 10 PW, both electric and magnetic fields get stabilized slightly below the threshold value, as shown in Fig. 2(a). At higher power, the plasma structure changes, as shown in Fig. 2(i-p) that leads to a slight deviation from the threshold. For lower powers the plasma distribution is well-localized inside the central part within the maximum of the electric field retaining sub-wavelength plasma size, but for higher powers a fraction of charged particles is pushed to the electric field minimum and this distribution becomes comparable to the wavelength. These particles have low energy, but their amount increases with power, which leads to a drop in average electron energy, see Fig. 2(f). At the same time, the maximum energy calculated by 1%, as shown in Fig. 1(j), follows the dependence of electric field on power and remains nearly constant and close to 1 GeV level for both positrons and photons, see Fig. 2(e). Averaged over the laser period plasma density in these structures reaches values up to $10^{25}$ cm$^{-3}$, (see Fig. 2(d)), which exceeds relativistic critical density. It should also be noted that current sheet distribution may be very narrow, as all particles attract to a single plane, and the maximum density values may be limited by grid resolution. The number of positrons $N_p$ in these sheets grows from $10^{10}$ for 8PW to $10^{11}$ for 15 PW. These particles oscillating along the z-axis create $J_z$ current, the total current in plane $z=0$ for different powers is plotted in Fig. 2(g). The maximum values of current and pair density are observed at the current instability stage and can be as high as 17 MA and $4.5\times10^{25}$ cm$^{-3}$ for 15 PW, respectively, see Fig. 2(d,g). During the development of instability the number of current sheets decreases, as well as the total number of particles, and the system relaxes to lower currents. In a stationary state for low power particles are mainly located within the central antinode region of electric field (radius $r < 0.44\ \lambda$). The current in this region increases with growth of power as the number of particles rises. For higher powers the particles are pushed to the next antinode region ($0.44\ \lambda < r < 0.97\lambda$) where the electric field as well as the generated current have opposite direction as compared to the central region, see Fig. 2(n), thereby decreasing the total current $J_z$ from 5 MA for 10 PW to approximately 3 MA for 15 PW, see Fig. 2(g).

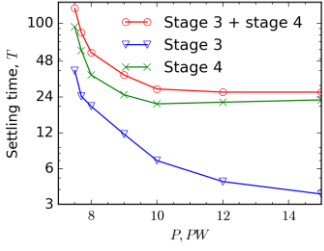

**Figure 3. Time scale of transition to stationary state.** Time for transition from the end of the linear stage to current instability stage – stage 3 in Fig. 1 (b) (blue line), the duration of the instability stage – stage 4 in Fig. 1 (b) (green line) and time for transition from the end of the linear stage to steady state – stages 3 and 4 in Fig. 1 (b) (red line).

Generated electrons and positrons have a significant probability to escape the focal region, for rough estimates we assume it to be 0.1. The electrons and positrons are emitted in the form of bursts with duration an order of $\tau_b \approx \frac{1}{4}T$ in opposite directions along the electric field ($z$-axis), so that in each direction the burst period equals T. Photons are emitted symmetrically in both directions along the $z$ axis with a period of $T/2$ with a $\pi/2$ phase shift relative to the particle bursts, see Fig. 1(i). The peak fluxes of both charged particles and photons with energies exceeding 1 GeV rise from a value of about $10^{23}$ for a laser power of 8 PW to $10^{24}$ for 15 PW as shown in Fig. 2(h). Based on the number of pairs in steady states, the total flux of pairs and photons can be estimated as $0.1\ N_p/\tau_b \approx 10^{25}$-$10^{26}$ s$^{-1}$ for power from 8 to 15 PW, which is in good correspondence with simulations. Correspondingly the total charge in one electron (positron) burst is about $0.1eN_p \approx$1-10 nC, however particles with energies exceeding 1GeV carry about tens of pC. Thus, by separating electrons and positrons in space, for example, by means of an external magnetic field, it is possible to control the total charge by choosing the duration of the laser pulse. For example, for the 30 fs duration of incident pulses a total charge of GeV particles can be up to 0.5 nC. Energy taken by particles from incident waves is much less than laser energy transformed into gamma radiation, see Fig 2(c). The conversion efficiency of laser energy into gamma photons approaches 55% for higher power and the absorption in high-density pair plasma may exceed 60%, shown in Fig. 2 (d).

**Control of the time scale**

Another important issue is achieving such extreme states of matter and antimatter in experiments. The characteristic time of transition to a stable steady state is shown in Fig. 3. For a power slightly above $P_{th}$, this time can exceed 100 optical cycles, which is quite natural, since the growth rate is low. For higher power it stabilizes at a level of about 25 laser cycles, which still exceeds the feasible duration of the laser pulse with a required power of 10-15 PW. This saturation can be explained based on the timeline shown in Fig. 1 (b). Laser-plasma dynamics passes through several stages, but the major part of this time is spent on the stage of the current sheets merger and relaxation to the final steady state of two aligned sheets (stage 4). This transition occurs under conditions which slightly depend on incident power, because at the instability stage electric field drops down to near-threshold level.

At first sight, such a long transition time looks non-optimistic on the way to realizing extreme states of matter and antimatter in experiments. We can argue this by two important considerations. First, the transient quasistationary stage of the current multi-sheet formation with the most extreme states can be achieved within 3-5 periods, even at not very high power of 10 PW. Second, development of current instability involves breaking of the axial symmetry, when the initially uniform distribution becomes a very narrow flat distribution. This process develops even for an ideal dipole wave without any explicit inhomogeneity, but initially non-uniform angular distribution of intensity or plasma can significantly accelerate the process of breaking symmetry and, thus, reaching a final steady state. This simple reason makes the multi-beam setup even more preferable, since an *n*-beam intensity distribution produces *n* seed sheets.

Another way to shorten the duration of transition to the stationary structure is control of plasma seed parameters. In order to attain extreme pair plasma states on a timescale of several optical cycles, we need to pass to the nonlinear stage of laser-plasma interaction as quickly as possible. For this it is necessary to consider targets with solid-state densities. First, such targets allow shortening the linear stage of the cascade. Second, heavier and slower ions reduce fast electron escape, decreasing $P_{th}$, therefore the cascade can develop during a larger part of the laser pulse. Third, dense targets can function as a plasma mirror (45). These targets have to reflect incident waves while instantaneous power is less than $P_{th}$. In this case the field penetrates into the plasma over the skin depth and does not cause abundant particle escape, thus plasma density should be about relativistic critical value $n_c\gamma$, where $\gamma$ corresponds to $P_{th}$. Since incident waves compress targets and ions lead to the decrease of the threshold of vacuum breakdown it is enough to consider initial densities of about $10^{22}$ cm$^{-3}$, according to PIC simulations.

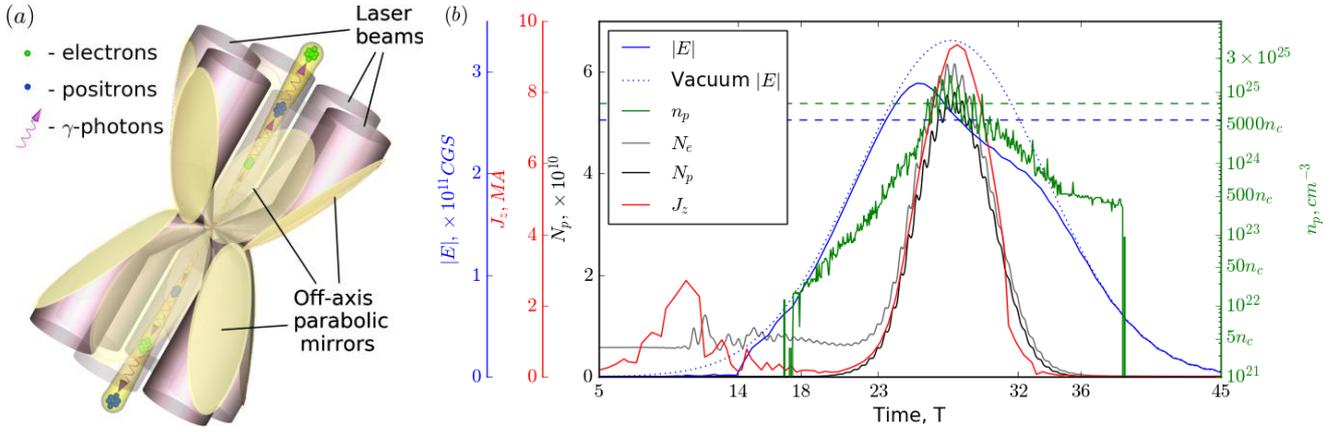

**Figure 4. Nonlinear interaction of a 30 fs 15 PW two-belt 12–beam lasers with dense plasma target.** (a) Schematic of proposed experimental configuration. (b) Timeline of interaction. Curves depict envelope of the electric field in the center of the simulation region $E$ (blue solid line), envelope of the vacuum electric field $E$ in the center of the simulation region (blue dotted line), field amplitude corresponding to stationary state, specifically, $P_{th}$ (blue dashed line), maximum positron density $n_p$ (green solid line), maximum electron-positron plasma density corresponding to stationary state (green dashed line), number of electrons in the cylinder with height and diameter equal to $\lambda$ $N_e$ (gray line), number of positrons in the same cylinder $N_p$ (black line), total current $J_z$ in through the z=0 plane (red line).

**Multi-beam setup**

We use aforementioned ideas to demonstrate the possibility of reaching extreme states at forthcoming laser facilities. In experiments the dipole-like wave structure can be mimicked by a number of tightly focused laser pulses. For a fixed power, the field distributions will be close to an ideal dipole wave but with lower maximum field intensity due to limited aperture compared to $4\pi$ dipole wave focusing. We suggest using the two-belt 12-beam laser system, a schematic of which is shown in Fig. 4(a). First of all, when the number of beams increases the total power needed to reach desired field intensity decreases, namely, to effectively simulate a 10 PW E-dipole wave we need 35 PW for a 4-beam configuration, 20 PW for a 6-beam configuration and only 12 PW for the proposed configuration using the same f-number optics. Thus, the required power in one beam is about 1 PW, which is already available on present-day laser facilities (46). Second, two-belt configuration allows forming a more optimal field structure in comparison with one-belt configurations since it covers a larger part of radiation pattern. Third, this setup still has a reasonable number of beams, taking into account experimental realization issues.

To be more specific about the properties of the extreme state achieved in such a complex geometry, we present the timeline of the nonlinear interaction in Fig. 4(b) based on 3D PIC modeling (see Methods section Numerical experiment setup). We would like to underline that the plasma-field dynamics passes through the same stages and the structure and properties of the achieved state are similar to those observed in a semi-infinite wave. In contrast to the considered low density case, at the compression stage a dense plasma target is not completely pushed to the center, and the field is mainly reflected. As soon as the field amplitude becomes high enough to penetrate into the dense plasma ($t\sim14T$), the electric field in the center grows unevenly and a structure close to the e-dipole wave is formed. Then, the cascade starts to develop leading to a rapid increase in the electron-positron plasma density. In this case, the initial density of the compressed electrons is sufficient to pass quickly to the nonlinear stage of the cascade development even at the rising edge of the pulse. Initially a nonuniform intensity distribution together with the high electron-positron plasma density lead to a very fast process of layering the current into a small number of sheets, which ultimately leads to the two-sheet distribution. The pair plasma density exceeds $10^{25}$ cm$^{-3}$, the value of the electric field is close to the breakdown threshold and photon flux up to $10^{23}$ photons with energies exceeding 1 GeV is generated during the interaction.

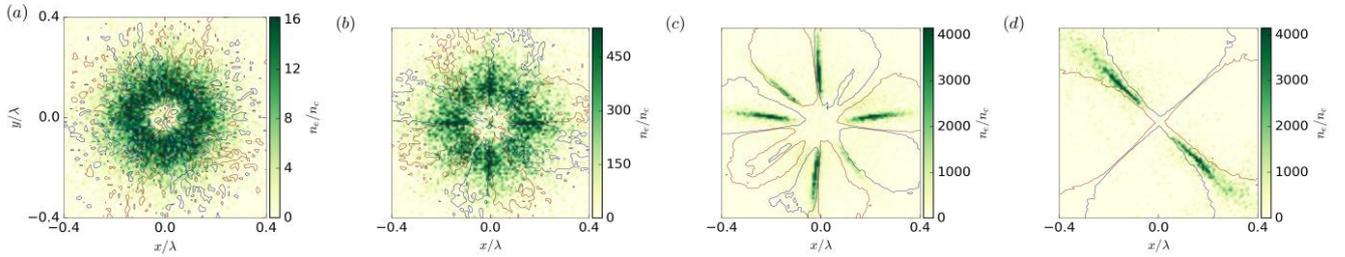

**Extended Data Figure 1. Symmetry breaking in the field of an e-dipole wave.** Snapshots of the plasma-field structures for a 10 PW e-dipole wave in the plane z=0. Contour field panel depicts $B\rho$ at 0.15 of maximum value, green color shows electron density normalized to the critical value for different moments of time: (a) 10T, uniform distribution; (b) 17.5T, multisheet stage; (c) 21.5T, nonlinear stage of four current sheets; (d) 46T, final stable distribution of two sheets. For better visualization magnetic field level curves are at the level 0.15 of the maximum value, maximum colorbar density value corresponds to 0.5 of the actual maximum value.

## Methods
### Threshold of the breakdown

Tight focusing of laser pulses leads to an increase of electric field strength and enhancement of electron-positron pair generation, however particles get the opportunity to escape the interaction focal region, thus reducing the growth rate or even leading to expansion of the seed target. For an e-dipole wave the main parameter is its power. If the power exceeds the petawatt level, a special ART regime emerges and traps particles. In this case they can leave the focal region mainly along the electric field, losing the opportunity to create pairs. The growth of power reduces particle escape due to trapping and intensifies pair creation. For a continuous wave, this ensures that there is a power threshold at which the particle losses are compensated by pair creation. To determine this threshold, we model the linear stage of the cascade in the dipole wave using the PICADOR PIC-code. As a seed, we consider a spherical plasma target with a radius of 1.5 wavelengths centered in the focus and a density of $10^{16}$ cm$^{-3}$ corresponding to 1000 real particles. Each timestep we count the number of pairs in the focal cylinder with a radius of 0.5 wavelength and height of 1 wavelength, which contains nearly all trapped particles. Their number averaged over a half-period of the wave grows or decreases exponentially. For near threshold powers of 6.5 PW and 8 PW, the total growth rates including pair creation and escape are -0.2/$T$ and 0.21/$T$, respectively. Linear interpolation shows that the approximate threshold is 7.2 PW.

### Numerical experiment setup

In this paper, we used two different numerical experiment setups. The first one is the semi-infinite wave with a rapid front to study characteristics of the extreme plasma states for a fixed wave power; the second setup employs Gaussian pulse shape with 30 fs duration to model a 12-beam experimental setup. In both setups laser radiation with a total power of 7.5-15 PW interacted with a plasma target with a 1.5μm radius. In the first case, the target density was changed in the range from $10^{15}$ to $10^{19}$ cm$^{-3}$ depending on power to ensure the presence of the linear stage of cascade. In the second case, the density was chosen so that the nonlinear stage is reached at the rear front of the pulse and corresponds to near solid densities of $10^{22}$ cm$^{-3}$. 12-beams with an f-number of 1.2 with a total power of 15 PW, which roughly corresponds to an ideal e-dipole wave of 13 PW, interacted with a plasma target with a radius of 1 wavelength and a density of $10^{22}$ cm$^{-3}$.

Numerical modeling was performed using Particle-in-Cell codes PICADOR (47) and ELMIS (48), which are based on FDTD and FFT electromagnetic solvers, respectively. In simulations the cubic grid with size of 4 μm x 4 μm x 4 μm and number of points of 512×512×512 was used. Due to the Courant stability criterion for the FDTD the time step in PICADOR code was chosen to be 0.01 fs. ELMIS code tolerates higher time steps, but we used the same values for consistency. We used the wavelength of 0.9 μm in conformity with the characteristics expected in the XCELS facility (3), but qualitatively the dynamics will be close for the setups with 0.8 μm wavelength. For a chosen wavelength spatial and temporal resolutions are approximately 115 and 300 steps per wavelength and period, respectively. For particle push the well-known Boris pusher (49) was used together with Esirkepov scheme (50) for current deposition. Special attention was paid to correct modeling of electron-positron cascade development, for this purpose a special Adaptive Event Generator module (29) was used with separate particle thinout for different particle types. This module has quite sophisticated algorithms, because it has to deal not only with correct modeling of pair and photon production, but these processes must be modeled correctly under conditions when density grows by several orders of magnitude within one period. This is achieved by using novel particles thinout techniques, when at a given step the particle ensemble is replaced by a smaller one with higher particle factors, and special procedures, including time step subdivision, allowing correct modeling of micro-avalanches within a single time step.

### Symmetry breaking

The most sudden effect in the formation of electron-positron-pair plasma is that even in the case of the highest symmetry of incident field configuration, plasma-field structures with essentially less symmetry are formed, as shown in Extended Data Fig. 1. This process of symmetry breaking can be explained only by the development of instability.

At the linear stage of cascade development plasma distribution is uniform and $B_\rho$ has a noise-like structure, see Extended Data Fig. 1(a). When plasma density approaches the relativistic critical density the distribution clearly becomes non-uniform, and the radial magnetic field component $B_\rho$ emerges together with density fluctuations, see Extended Data Fig. 1(b). Since co-directional currents are attracted to each other, at the nonlinear stage of instability the adjacent sheets merge along the azimuth, while radial merging is suppressed by the ponderomotive force. A highly inhomogeneous current structure is formed with several distinct well-seen current sheets; in this case the characteristic $B_\rho$ value is of the order of 0.1 of the vacuum field, as in Extended Data Fig. 1 (c). Although the field structure is partly modified due to dense current sheets, particle motion is still ART-like and mainly in the plane. Finally, the stable configuration of two sheets at the angle of $\pi$ from each other is formed; particles during their motion do not cross the $z$ axis, as is shown in Extended Data Fig. 1(d). In this case the value of $B_\rho$ becomes comparable with the value of the vacuum field. Plasma tends to gather at the nodes of $B_\rho$, leading to extremely high densities, that computer simulations do not properly resolve. Indeed, doubling the number of cells along the $x$ and $y$ directions of the simulation box leads to doubled maximum density. It is also interesting to note that the final current direction is not the most prominent one at intermediate stages, which indicates a complex nature of the current dynamics.

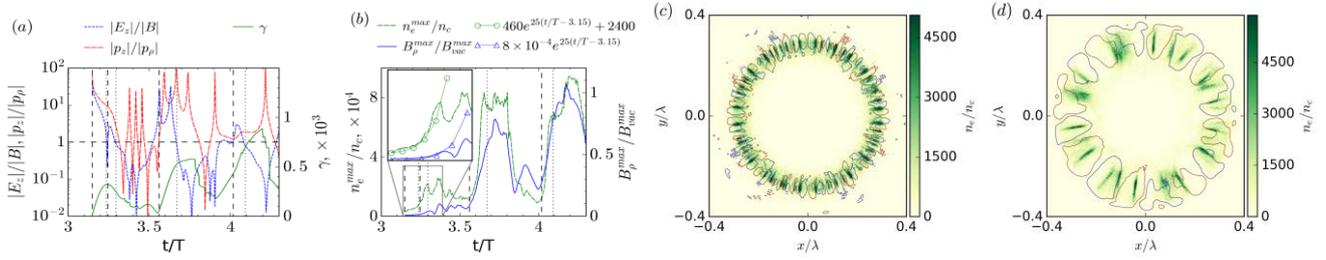

**Extended Data Figure 2. Current instability of electron-positron plasma in the field of cylindrical wave.** Amplitude of electric field in focus is $a = 2500$. (a) Ratio of electric field to magnetic field (dashed line), ratio of $p_z$ to radial electron momentum (dash-dotted line), and gamma factor (solid line) along typical particle trajectory. (b) Time evolution of maximum electron density and magnitude of radial magnetic field during development of instability. Horizontal dashed line represents unity level. The time interval when electric field and $p_z$ exceed magnetic field and $p_\rho$ respectively is between vertical dashed and dotted lines. In the inset the exponential approximation is shown by lines with symbols. Spatial distribution of generated radial component of the magnetic field (colored level curves) and electron density at time moments (c) $t = 3.48T$, (d) $t = 3.95T$. For better visualization level curves are at the level $\pm \frac{1}{7} B_\rho^{max}$.

A very intriguing question is: "What are the mechanisms of this instability that violate the symmetry of the laser-plasma interaction?" In general, they can have an electrodynamic nature, as the plasma interacts strongly with laser field. Paying attention to the fact that the instability is small-scaled with density stratifications along the electric field, and the entire plasma structures are well localized in the focal volume with dimensions smaller than the laser wavelength, we conclude that this is not the result of nonlinear laser plasma interactions themselves. Moreover, since the pair plasma is the streams of electrons and positrons moving in opposite directions in the electromagnetic field which acts as a driver, current instability can arise even in a given field.

**Cylindrical e-type wave.** To get an insight into the physical mechanisms causing the instability, we consider a simple two-dimensional case: the interaction of an e-type cylindrical wave with plasma distributed within a cylindrical ring. This setup allows us to analyze the pure axisymmetric case with uniformity along the $z$ axis. In this simplistic case, there are only $E_z$ and $B_\varphi$ field components, but the field structure is very close to the field distribution in the central cross-section $z = 0$ of the dipole wave.

We performed a PIC simulation for the following parameters. Standing cylindrical wave with amplitude $a=2500$ in relativistic units interacts with electron-positron plasma $n_0=2400n_c$ located within $0.17\lambda<\rho<0.27\lambda$, which correlates with the range of radial oscillation in ART regime. The chosen density value corresponds to the beginning of the current instability stage in the 10 PW dipole wave. The initial moment of the interaction is when the magnetic field is zero at the time $t = 3.15T$. For simplicity, pair generation is switched off. The result of the PIC simulation is shown in Extended Data Fig. 2.

Initially, the particles are accelerated along the electric field, increasing mainly the longitudinal component of the momentum and reach the maximum gamma factor $\gamma_{max}=300$, see Extended Data Fig. 2(a). Along with this the azimuthally modulated plasma-field structure emerges from the noise. The initial perturbation of radial magnetic field and density are equal to $B_\rho^0 \approx 8 \times 10^{-4} B_{max}^{vac}$ and $\tilde{n}_0 \approx 460 n_c$. They grow fast so that $0.15T$ from the beginning of the interaction, the maximum density becomes an order of magnitude greater than the initial density, and the radial component of the magnetic field becomes 0.1 of the maximum azimuthal component of the standing wave $B_{max}^{vac}$, see insert in Extended Data Fig. 2 (b). Spatial plasma-field distributions show the same behavior as in the dipole wave. Initially, many current sheets are generated, but later they merge, see Extended Data Fig. 2 (c,d) and radial and azimuth magnetic fields become comparable. Detailed comparison of Extended Data Figs. 2(a,b) shows that the perturbations mainly develop when the motion of the particles proceeds along the $z$ axis and the axial electric field exceeds the azimuthal magnetic field. For these time intervals taking into account rapid development of instability, we can derive a simple model of the instability based on hydrodynamics equations coupled with the Maxwell equations.

**The instability model.** Assuming that the growth rate is much greater than the field frequency, we suppose that the distributions of the plasma and the electromagnetic field are approximately constant in time. Within the considered time interval electrons and positrons move in opposite directions mainly along the $z$ axis with velocity $v_0$ and the radial motions are negligible, the corresponding gamma factor is $\gamma_0$. According to Extended Data Fig. 2(a), during these periods of time the azimuthal magnetic field can be neglected, thus we omit the radiation reaction force. Validity of this assumption is confirmed by the absence of significant energy leaps due to the photon emission during these time intervals, see green solid line in Extended Data Fig. 2(a). Since the radial motion of the plasma is limited by ART in a relativistically strong field, it is natural to expect that the perturbations will have the form of $e^{\Gamma t + il\varphi}$, where $\Gamma$ is the instability growth rate, $l$ is an integer and $\varphi$ is the azimuthal angle. From the Maxwell equations it follows that perturbation of plasma density $\tilde{n}$ lead to generation of the radial component of the magnetic field $B_\rho = \dfrac{i 8\pi e \rho v_0 \tilde{n}}{lc}$. Here we neglect perturbations of axial electric and azimuthal magnetic fields. This assumption needs $l \gg \Gamma/\omega_l$. The applicability of the assumption is confirmed by PIC-code simulation since initially many sheets are created, see Extended Data Fig. 2(c). At the same time from the hydrodynamics equations: $\Gamma \tilde{n} + i \dfrac{l}{\rho} n_0 v_\varphi = 0$ and $\Gamma m v_\varphi \gamma_0 = \dfrac{e}{c} v_0 B_\rho$, we see that $B_\rho$ excites azimuthal motion with velocity $v_\varphi$. Pairs are attracted to the nodes of $B_\rho$, and perturbations of density increase as $\tilde{n} = \dfrac{-ilB_\rho e n_0 v_0}{\Gamma^2 \rho m \gamma_0 c}$. Thus the self-excited process of the current instability occurs with the growth rate $\Gamma = \dfrac{\omega_p}{\sqrt{\gamma_0}} \dfrac{v_0}{c} \cong \dfrac{\omega_p}{\sqrt{\gamma_0}}$, where $\omega_p = \sqrt{\dfrac{8\pi e^2 n_0}{m}}$ is the plasma frequency.

For the realistic parameters (see **Cylindrical e-type wave**) of $n_0=2400n_c$, $\gamma_0 \approx \gamma_{max}/2 = 150$ the estimate of the instability growth rate *Γ≈25/T* corresponds quite closely to the results of numerical simulations shown in Extended Data Fig. 2 (b). The following important consequences confirm the derived model of the instability. First, the increment does not depend on the wave number of the disturbance, which indicates the possibility of growth of perturbations of different scales, which is actually reflected in Extended Data Figs. 1(b) and 2(c). Second, due to fixed radial particle motion in the laser field plasma decomposes into sheets along the azimuth angle, not filaments. Third, the increment is anomalously large and formally close to the growth rate of the Weibel instability (51), but unlike the latter, the current sheets are purely quasi-neutral, which completely agrees with the simulations.

**Author Contributions**
All authors contributed at all stages of the work presented here.

**Acknowledgements**
The authors acknowledge support from the Russian Science Foundation project No. 16-12-10486.